\documentclass[preprint,12pt]{elsarticle}

\usepackage[OT1]{fontenc}
\usepackage{amsthm,amsmath}
\usepackage[numbers]{natbib}
\usepackage[colorlinks,citecolor=blue,urlcolor=blue]{hyperref}

\usepackage[utf8]{inputenc}
\usepackage{graphicx}
\usepackage{subcaption}
\usepackage{algorithm,algorithmic}
\usepackage{multirow}
\usepackage{array}

\usepackage{amssymb}

\begin{document}

\begin{frontmatter}



\title{Structural Equation Modeling and simultaneous clustering through the Partial Least Squares algorithm}


\author[label1]{Mario Fordellone}
\author[label1]{Maurizio Vichi}

\address[label1]{Sapienza, University of Rome}

\begin{abstract}
The identification of different homogeneous groups of observations and their appropriate analysis in PLS-SEM has become a critical issue in many application fields. Usually, both SEM and PLS-SEM assume the homogeneity of all units on which the model is estimated, and approaches of segmentation present in literature, consist in estimating separate models for each segments of statistical units, which have been obtained either by assigning the units to segments \textit{a priori} defined. However, these approaches are not fully acceptable because no causal structure among the variables is postulated. In other words, a modeling approach should be used, where the obtained clusters are homogeneous with respect to the structural causal relationships.

In this paper, a new methodology for simultaneous non-hierarchical clustering and PLS-SEM is proposed. This methodology is motivated by the fact that the sequential approach of applying first SEM or PLS-SEM and second the clustering algorithm such as $K$-means on the latent scores of the SEM/PLS-SEM may fail to find the correct clustering structure existing in the data. A simulation study and an application on real data are included to evaluate the performance of the proposed methodology.

\end{abstract}


\end{frontmatter}


\section{Introduction}
In the last years, structural equation modeling (SEM) has become one of the reference statistical methodologies in the analysis of the statistical relationships between observable (manifest) and non-observable (latent) variables. SEM are often used for both to assess non-observable \textit{hidden} constructs (i.e., latent variables) by means of observed variables, and to evaluate the relations among latent constructs and among manifest variables. In SEM, variables (manifest or latent) are considered $(i)$ endogenous if they are dependent, i.e., related to a set of variables that explain them; $(ii)$ exogenous if they are independent, i.e., explain a set of variables.. Note that endogenous variables may also cause other endogenous variables. SEM has the property to estimate the multiple and interrelated dependencies in a single analysis by combining factor analysis and multivariate regression analysis. SEM has been used in many different fields, as in economics and social sciences, in marketing for example to assess customer satisfaction (Fornell and Larcker, 1981; Ringle et al., 2012; Steenkamp and Baumgartner, 2000; Squillacciotti, 2010). Then, SEM allows to build latent variables (LVs), such as customer satisfaction, through a network of manifest variables (MVs). 

Covariance structure approach (CSA) (Jöreskog, 1978) and partial least squares (PLS) (Lohmöller, 1989) are the two alternative statistical techniques for estimating such models. The CSA, also referred to the most well-known LISREL model, uses the ML estimation; thus, has the advantage to allow the researcher to make inference on the results. However, PLS is considered desirable in three specific cases: $(i)$ when the sample size is small, $(ii)$ when the data to be analyzed is not multinormal as required by CSA, and $(iii)$ when the complexity of the model to be estimated may lead to improper or non-convergent results (Bagozzi and Yi, 1994; Squillacciotti, 2010). 

In CSA the parameter estimation is obtained by comparing the covariance matrix of the manifest variables with the covariance matrix derived by the structural and measurement models, and optimizing a discrepancy function of this comparison. PLS is an iterative estimation method introduced by the Swedish statistician Herman Wold and his son Svante Wold. It is also called projection to latent structures (Wold et al., 2001), but the term PLS is still dominant in many areas. Although PLS is widely used in many fields, within the realm of psychology, it has received criticism for being considered is some situations an unreliable estimation and testing tool (Rönkkö et al., 2016).

An important research objective that we wish to study in this paper is the simultaneous classification of the statistical units together with the structural equation modeling of the manifest variables. For example, a relevant issue in marketing is the measurement of customer satisfaction by using SEM or PLS-SEM. However, also the identification of distinctive customer segments (Hofstede et al., 1999; Wu and Desarbo, 2005) has been considered relevant together with these methodologies. In fact, the identification of different groups (clusters) of observations and their appropriate analysis in PLS-SEM has become a critical issue in many application fields. This because, usually, SEM implicitly assumes via the multi-normality distribution of the units their homogeneity. However, in many cases, this assumption may turn out to be imprecise and the presence of latent classes not accounted for by the global model may lead to biased or erroneous results in terms of parameters estimation and model quality (Ringle et al., 2012; Squillacciotti, 2010).

In literature, different models exist that allow a reliable identification of distinctive segments of observations in the SEM context (Sörböm, 1974). The majority of these models consist in estimating separate models for each homogeneous cluster, which has been obtained either by \textit{a priori} assignment based, e.g., on demographic variables, or through a cluster analysis applied on the original variables. However, this sequential approach based on the application of clustering and SEM on each cluster may fail because the preventive clustering is not functional to define the best successive structural and measurement relations among variables. In other words, a simultaneous estimation of clusters and the structural/measurement causal relationships should be considered, so that the obtained clusters are the most homogeneous that best explain the relationships between variables.  

In this direction Jedidi et al. (1997) propose a simultaneous procedure based on finite mixture estimated via the expectation-maximization (EM) algorithm (Dempster et al., 1977; McLachlan and Krishnan, 2004; Wedel and Kamakura, 2002). Hahn et al. (2002) affirm that this technique extends CSA, but it is inappropriate for PLS-SEM.  They propose the finite mixture partial least squares (FIMIX-PLS) approach that joins a finite mixture procedure with an EM algorithm specifically regarding the ordinary least predictions of PLS. Sarstedt (2008) reviews this technique and concludes that FIMIX-PLS can currently be viewed as the most comprehensive and commonly used approach for capturing heterogeneity in PLS-SEM. Following the guidelines of Jedidi et al. (1997) and Hahn et al. (2002), Ringle et al. 2005 present FIMIX-PLS implemented for the first time in a statistical software application, called Smart-PLS. Vinzi et al. (2008) propose a new method for unobserved heterogeneity detection in PLS-SEM: response-based procedure for detecting unit segments in PLS (REBUS-PLS). REBUS-PLS does not require distributional hypotheses but may lead to local models that are different in terms of both structural and measurement models. In fact, separate PLS-SEM are estimated for each cluster, and the results are compared in order to identify, if possible, differences among component scores, structural coefficients and different loadings. This is certainly an interesting feature, which has the unique problem of complicating the interpretation of results, since the number of the SEM parameters to be mentioned increases at the increasing of the number of clusters. Following this idea, Squillacciotti (2010) proposes a technique, called PLS typological path modeling (PLS-TPM), that allows to take into account the predictive purpose of PLS techniques when the classes are defined. 

Otherwise, the researcher could consider the sequential approach of applying first SEM or PLS-SEM in order to determine the LVs and then apply a clustering methodology such as $K$-means or Gaussian mixture model (GMM) clustering on the latent scores of the SEM/PLS-SEM in order to obtain homogeneous clusters. However, Sarstedt and Ringle (2010) empirically illustrate the shortcomings of using a sequential approach. For this reason, other more reliable methodologies have been introduced in the last years. These include prediction oriented segmentation in PLS path models (PLS-POS) proposed by Becker et al. (2013), genetic algorithm segmentation in partial least squares path modeling (PLS-GAS) proposed by Ringle et al., (2013), and particularly segmentation of PLS path models through iterative reweighted regressions (PLS-IRRS) proposed by Schlittgen et al. (2016). For more details see also Sarstedt et al. (2017). Schlittgen et al. (2016) conclude that PLS-IRRS gives similar quality results in comparison with PLS-GAS, and it is generally applicable to all kinds of PLS path models. Moreover, the PLS- IRRS computations are extremely fast. 

Note that in all the segmentation methods discussed above, researchers must pre-specify a number of segments (clusters) when running the procedure. The optimal number of segments is usually unknown. Ringle et al. (2013) and Schlittgen et al. (2016) propose to firstly run FIMIX-PLS  (Hahn et al., 2002; Sarstedt and Ringle, 2010) to determine the number of segments and, then, subsequently run PLS-GAS or PLS-IRRS to obtain the final segmentation solution.

Following other research fields, also Vichi and Kiers (2001) warn against the use of the “tandem analysis”, i.e., the sequential application of a dimensionality reduction methodology (e.g., principal component analysis or factor analysis) and then a clustering algorithm (e.g., $K$-means or mixture models) on the factor scores, because this application may fail to identify the clustering structure existing in the data. In fact, the factorial methodology identifies LVs that generally explain the largest variance of the MVs; while the clustering methodology tends to explain the between variance only. Thus, if the data include MVs that do not have the clustering structure, the factorial methodology will explain also these last (especially if they have a large variance), and thus tandem analysis will mask the subsequent estimation of the clustering structure in the data. Now, we show that the sequential approach of PLS-SEM and $K$-means may fail to find the clustering structure in the data, as well as in the cases described in Vichi and Kiers (2001). 

\begin{figure}[!h]
\begin{subfigure}{.33\textwidth}
  \centering
  \includegraphics[width=1\linewidth]{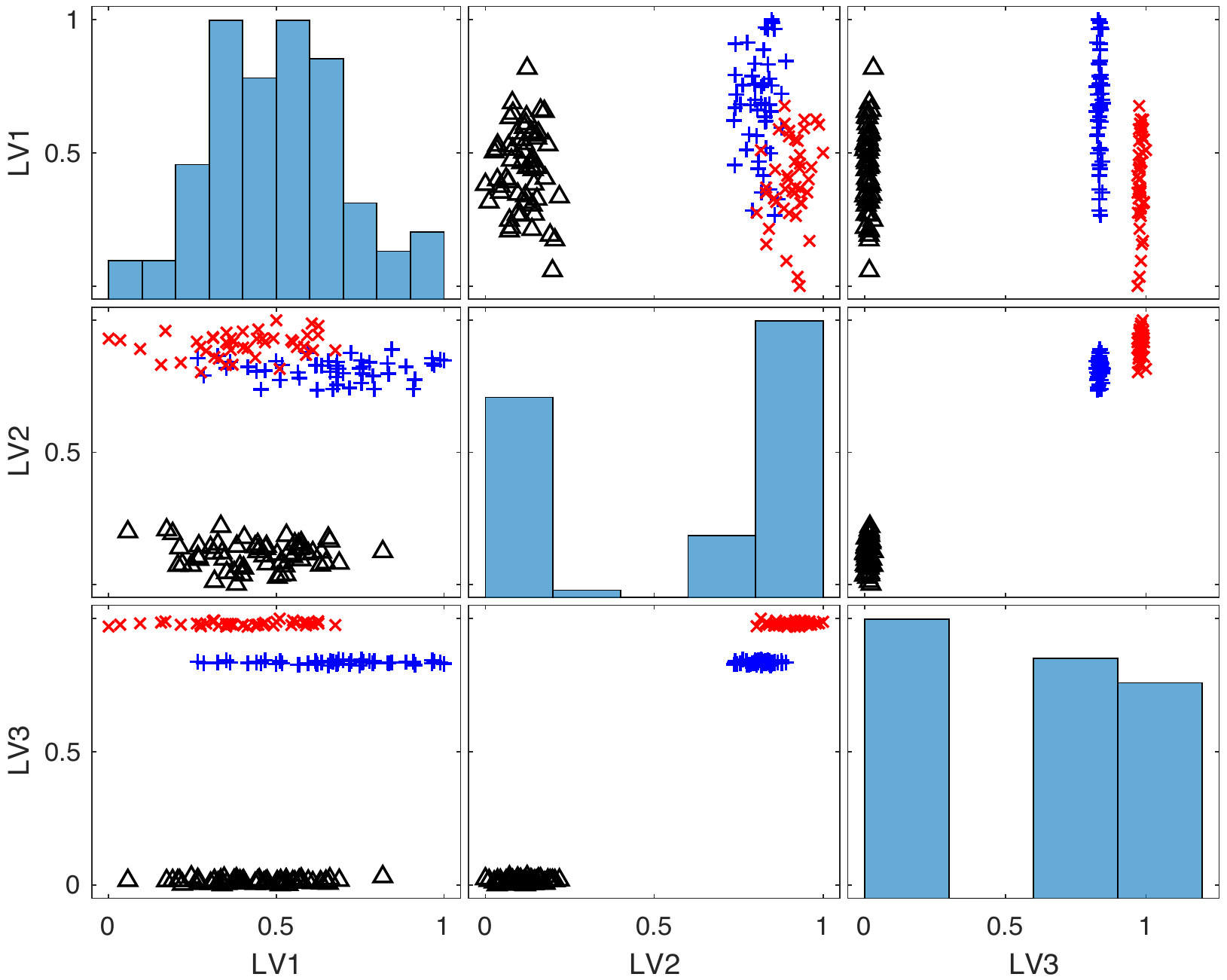}
  \caption{}
  \label{fig:sfig1}
\end{subfigure}%
\begin{subfigure}{.33\textwidth}
  \centering
  \includegraphics[width=1\linewidth]{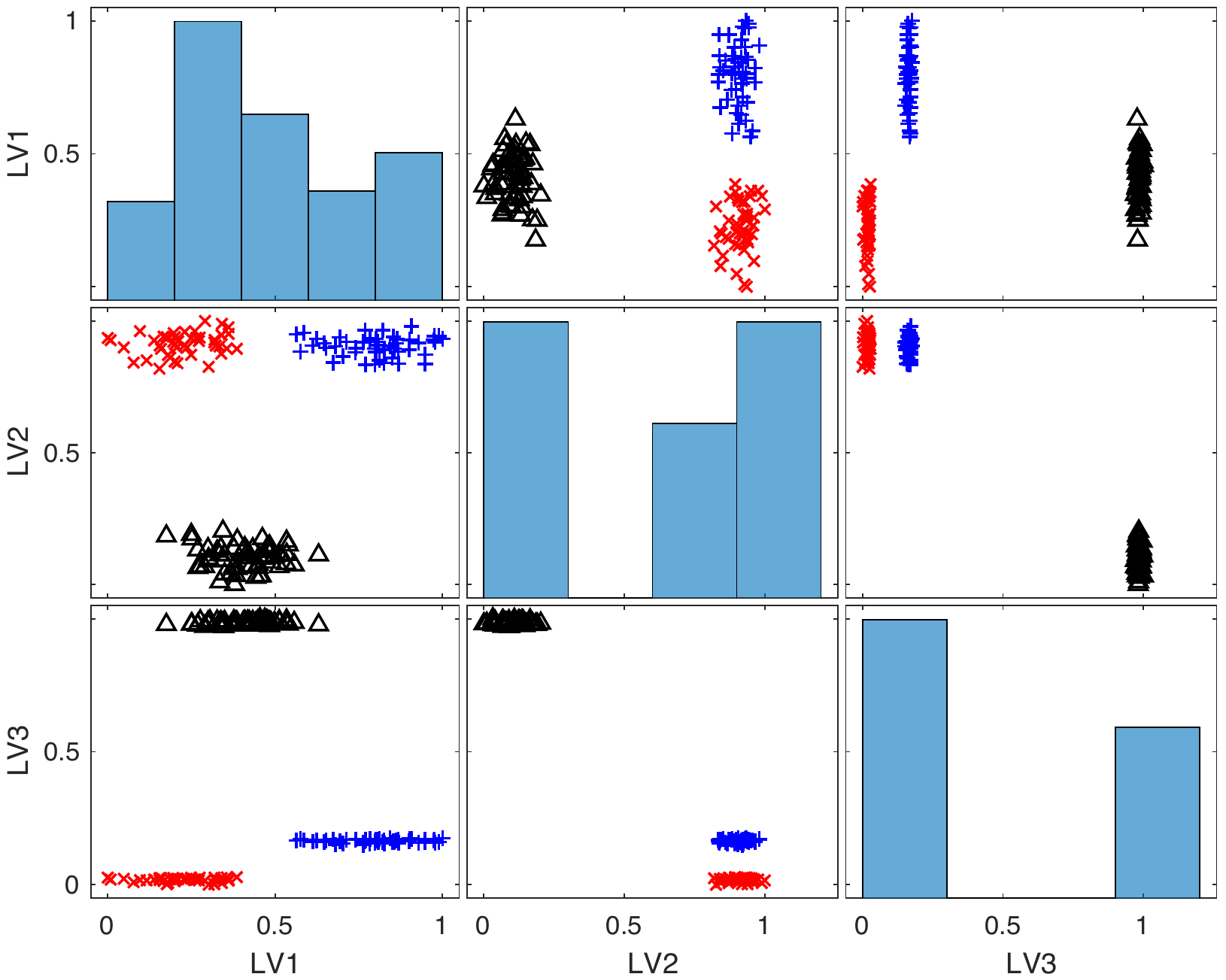}
  \caption{}
  \label{fig:sfig2}
\end{subfigure}
\begin{subfigure}{.33\textwidth}
  \centering
  \includegraphics[width=1\linewidth]{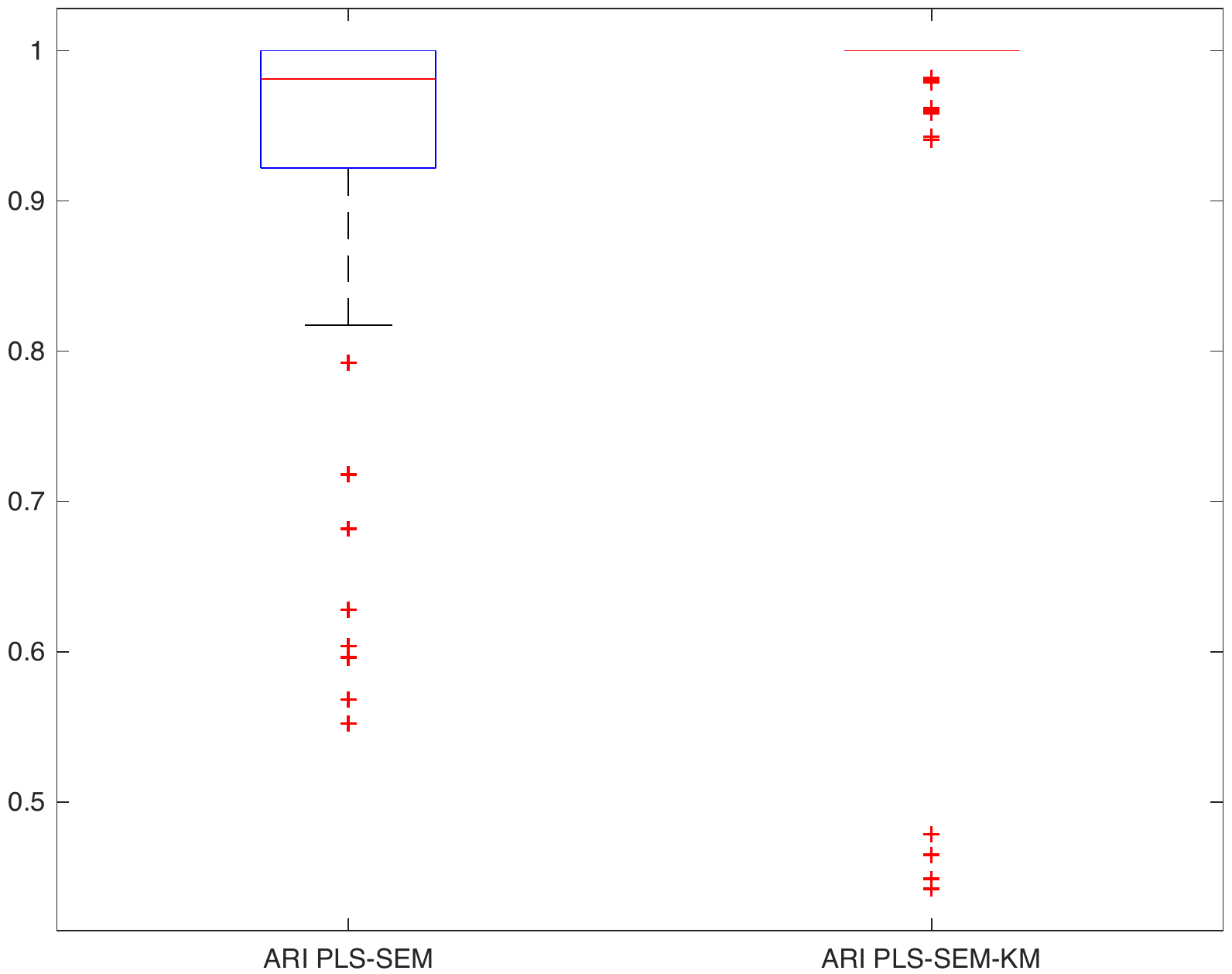}
  \caption{}
  \label{fig:sfig3}
\end{subfigure}
\caption{Left figure represents the scatterplot-matrix of the LVs estimated by the sequential application of PLS-SEM and $K$-means; center figure represents the scatterplot-matrix of the LVs estimated by the simultaneous application of PLS-SEM and $K$-means; right figure represents the boxplot of ARI distribution between the true and estimated partition obtained by the sequential and simultaneous approaches on 300 data sets.}
\end{figure}

The data set on which we apply sequentially PLS-SEM and $K$-means is formed as follows. Two exogenous LVs, having a clustering structure into three groups, have been generated by a mixture of three circular bivariate normal distributions, with mean vectors located at the vertices of a equilateral triangle. Then, an endogenous LV has been generated by a normal distribution with $\mu=0$ and $\sigma^2=3$. Each LV reconstructs three MVs. Together with these $3\times 3$ reconstructed MVs we have also added other two MV, for each of the 3 LVs. These new MVs do not have clustering structure, since they are generated by a single circular bivariate normal distribution with $\mathbf{\mu=\underline{0}}$ and $\mathbf{\Sigma=6I}$. Thus, the data set is formed by 9 MVs with clustering structure (6 directly generated and 3 induced by the relation between exogenous and endogenous variables) and 6 noise variables that tend to mask the groups in the data. In Figure \ref{fig:sfig1} the scatterplot matrix of the three LVs shows the performance of the sequential application of PLS-SEM and $K$-Means. The clusters are not well separated; in fact, the adjusted Rand index (ARI) (Hubert and Arabie, 1985) between the generated partition and the partition obtained by $K$-means computed on the LVs of the PLS-SEM is equal to 0.64. Note that ARI is equal to 0 when two random partitions are compared and it is equal to 1 when two identical partitions are compared.
In Figure \ref{fig:sfig2} the scatterplot matrix of the LVs shows the simultaneous application of PLS-SEM and $K$-means as proposed in this paper with a specific methodology. This time the clusters are well separated and the partition is clearly detected. ARI is equal to 1, i.e., the new methodology exactly identifies the generated partition. We might think that this is only a simulated example. Thus, we have repeated the analysis on other 300 data sets generated as above described. The boxplot of the distribution of the 300 adjusted Rand index (ARI) (Figure \ref{fig:sfig3}) confirms that,  in almost all cases the new methodology exactly finds the generated partition (mean of ARI equal to 0.98), while the sequential application of PLS-SEM and $K$-means finds the true partition in 15\% of cases and the mean of ARI is equal to 0.65.    

Thus, we can conclude that the sequential application of PLS-SEM and $K$-means is an unreliable approach to obtain the best partition of the units and the best structural equation modeling analysis when data are heterogeneous. This result strongly motivates the study of a new model that identifies simultaneously the best clustering and the best manifest variables reconstructed by a unique common set of measurement/structural relationships. The new methodology is estimated with PLS and therefore is named partial least squares $K$-means (PLS-SEM-KM). The model is based on the simultaneous optimization of PLS-SEM and reduced $K$-means (De Soete and Carroll, 1994), where centroids of clusters are located in the reduced space of the LVs, thus, ensuring the optimal partition of the statistical units on the best latent hyper-plane defined by the structural/measurement relations estimated by the pre-specified model. Moreover, a different approach to select the optimal number of segments is provided.

The paper is structured as follows. In the section 2 a brief background on the SEM estimated via PLS procedure in provided. In section 3 the PLS-SEM-KM model is presented and the PLS algorithm is given. In section 4 the performances of PLS-SEM-KM are tested in a detailed simulation study providing a comparison with the FIMIX-PLS approach proposed by Hahn et al. (2002). In section 5 the results obtained by an application on real data are shown.

\section{Notation and Partial Least Squares in Structural Equation Modeling}
\subsection{Matrix notation}
Before showing the modeling details, the notation and terminology used in this paper is here presented to allow the reader to easily follow the subsequent formalizations and algebraic elaborations.
\begin{flushleft}
\begin{tabular}{| l l l |}
\hline
$n$, $J$ & \# of: & observations, MVs\\ 
$H$, $L$, $P$  & \# of: & exogenous LVs, endogenous LVs, LVs ($P=H+L$)\\
$K$  & \# of: & clusters\\
$\mathbf{\Xi}$   &  $n\times H$   &  exogenous LVs matrix\\
$\mathbf{H}$    &  $n\times L$    &  endogenous LVs matrix\\
$\mathbf{Y}$    &  $n\times P$     &  scores matrix (${\mathbf{Y}=[\mathbf{\Xi},\mathbf{H}]}$)\\
$\mathbf{\Gamma}$  &  $L\times H$       &   path coefficients matrix of the exogenous LVs\\
$\mathbf{B}$    &  $L\times L$     &  path coefficients matrix of the endogenous LVs\\
$\mathbf{Z}$  &  $n\times L$ & errors matrix of the endogenous LVs\\
$\mathbf{X}$    &  $n\times J$    &  data matrix\\
$\mathbf{E}$  &  $n\times J$  & errors matrix of the data\\
$\mathbf{\Lambda}_H$  &  $J\times H$  & loadings matrix of the exogenous LVs\\
$\mathbf{\Lambda}_L$  &  $J\times L$  &  loadings matrix of the endogenous LVs\\
$\mathbf{\Lambda}$  &  $J\times P$ &loadings matrix (${\mathbf{\Lambda}=[\mathbf{\Lambda}_H,\mathbf{\Lambda}_L]}$)\\
$\mathbf{T}$  &  $n\times H$  & errors matrix of the exogenous LVs\\
$\mathbf{\Delta}$  &  $n\times L$  &errors matrix of the endogenous LVs\\
\hline
\end{tabular}
\end{flushleft}

Partial least squares (PLS) methodologies are algorithmic tools with analytic proprieties aiming at solving problems about the stringent assumptions on data, e.g., distributional assumptions that are hard to meet in real life (Tenenhaus et al., 2005). Tenenhaus et al. (2005) try to better clarify the terminology used in the PLS field through an interesting review of the literature, focusing the attention on the structural equation models (SEM) standpoint.\\
Usually, a PLS-SEM (called also PLS-PM, i.e., PLS path model) consists in a combination of two models: 
\begin{itemize}
    \item a \textit{structural model} (or inner model), that specifies the relationships between latent variables (LV). In this context, a LV is an non-observable variable (i.e., a theoretical construct) indirectly described by a block of observable variables which are called manifest variables (MVs);
    \item a \textit{measurement model} (or outer model), that relates the MVs to their own LVs.
\end{itemize}
\subsection{Structural model}
\begin{figure}[!h]
\centering
\includegraphics[scale=0.5]{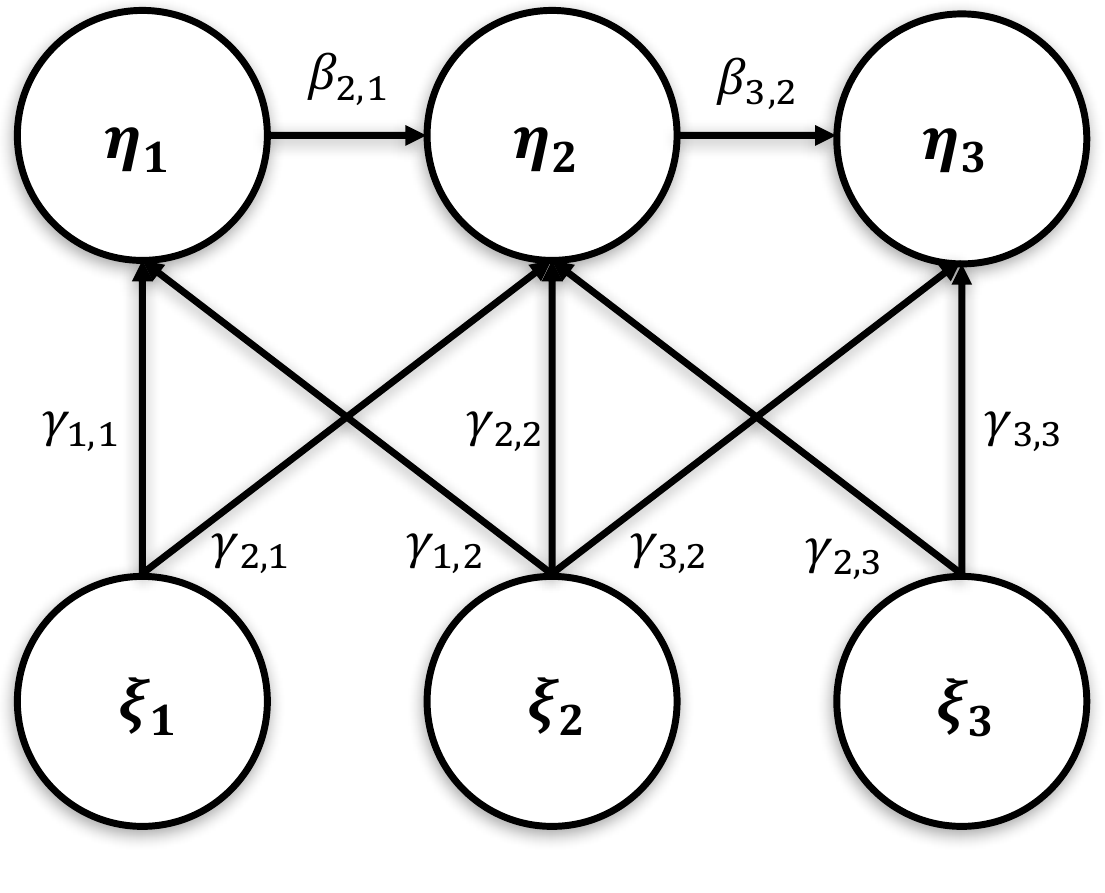}
\caption{Example of structural model with three endogenous LVs and three exogenous LVs}\label{fig:2}
\end{figure}

Let $\mathbf{X}$ be a $n\times J$ data matrix, summarized by $P$ latent variables ($j=1,\dots, j=J$; $p=1,\dots, p=P$ and $P\leq J$), let $\mathbf{H}$ be the $n\times L$ matrix of the endogenous LVs with generic element $\eta_{i,l}$, and let $\mathbf{\Xi}$ be the $n\times H$ matrix of the exogenous LVs with generic element $\xi_{i,h}$, the structural model is a causality model that relates the $P$ LVs each other through a set of linear equations (Vinzi et al., 2010). In matrix form:
\begin{equation}\label{eq:1}
    \mathbf{H}=\mathbf{HB}^T+\mathbf{\Xi\Gamma}^T+\mathbf{Z},
\end{equation}
where $\mathbf{B}$ is the $L\times L$ matrix of the path coefficients $\beta_{l,l}$ associated to the endogenous latent variables, $\mathbf{\Gamma}$ is the $L\times H$ matrix of the path coefficients $\gamma_{l,h}$ associated to the exogenous latent variables, and $\mathbf{Z}$ is the $n\times L$ matrix of the residual terms $\zeta_{i,l}$.\\
\underline{\textbf{EXAMPLE 1}}. An example of structural model is shown in Figure \ref{fig:2}. Three endogenous LVs and three exogenous LVs are reported.  
\\The structural equations related to the path diagram in Figure \ref{fig:2} are shown in compact matrix form in Equation (\ref{eq:2}).
\begin{equation}\label{eq:2}
    \begin{bmatrix}\eta_1\\ \eta_2\\ \eta_3\\ \end{bmatrix}^T=
\begin{bmatrix}\eta_1 \eta_2 \eta_3 \end{bmatrix}
\begin{bmatrix}
0 &0  &0 \\ 
\beta_{2,1} &0  &0 \\ 
0 &\beta_{3,2}  &0 
\end{bmatrix}^T+
\begin{bmatrix}\xi_1 \xi_2 \xi_3 \end{bmatrix}
\begin{bmatrix}
\gamma_{1,1} &\gamma_{1,2}  &0 \\ 
\gamma_{2,1} &\gamma_{2,2}  &\gamma_{2,3} \\ 
0 &\gamma_{3,2}  &\gamma_{3,3} 
\end{bmatrix}^T+
\begin{bmatrix}\zeta_1\\ \zeta_2\\ \zeta_3\\ \end{bmatrix}^T.
\end{equation}
\subsection{Measurement model}
In PLS-SEM, unlike traditional SEM approach, there are two ways to relate MVs to their LVs: \textit{reflective} way and \textit{formative} way (Diamantopoulos and Winklhofer, 2001; Tenenhaus et al., 2005).
In the reflective way it is supposed that each MV reflects its LV, i.e., the observed variables are considered as the effect of the latent construct; a reflective measurement model can be written in matrix form as
\begin{equation}\label{eq:3}
\begin{split}
    \mathbf{X}&=\mathbf{Y\Lambda}^T+\mathbf{E}\\
              &=\begin{bmatrix}\mathbf{\Xi} & \mathbf{H}\end{bmatrix} \begin{bmatrix} \mathbf{\Lambda}_H^T\\ \mathbf{\Lambda}_L^T  \end{bmatrix}+\mathbf{E}\\
              &=\mathbf{\Xi\Lambda}_H^T+\mathbf{H\Lambda}_L^T+\mathbf{E},
\end{split}
\end{equation}
where $\mathbf{\Lambda}_H$ is the $J\times H$ loadings matrix of the exogenous latent constructs with generic element $\lambda_{j,h}$, $\mathbf{\Lambda}_L$ is the $J\times L$ loadings matrix of the endogenous latent constructs with generic element $\lambda_{j,l}$, and $\mathbf{E}$ is the $n\times J$ residuals matrix with element $\epsilon_{i,j}$, under hypothesis of zero mean and is uncorrelated with $\xi_{i,h}$ and $\eta_{i,l}$. Then, the reflective way implies that each MV is related to its LV by a set of simple regression models with coefficients $\lambda_{j,l}$. Note that, in reflective way is necessary that the block of MVs is unidimensional with respect to related LV. This condition can be checked through different tools, as well as principal component analysis (PCA), Cronbach’s alpha and Dillon-Goldstein’s  (Vinzi et al., 2010; Sanchez, 2013).  

\begin{figure}[!h]
\centering
\includegraphics[scale=0.5]{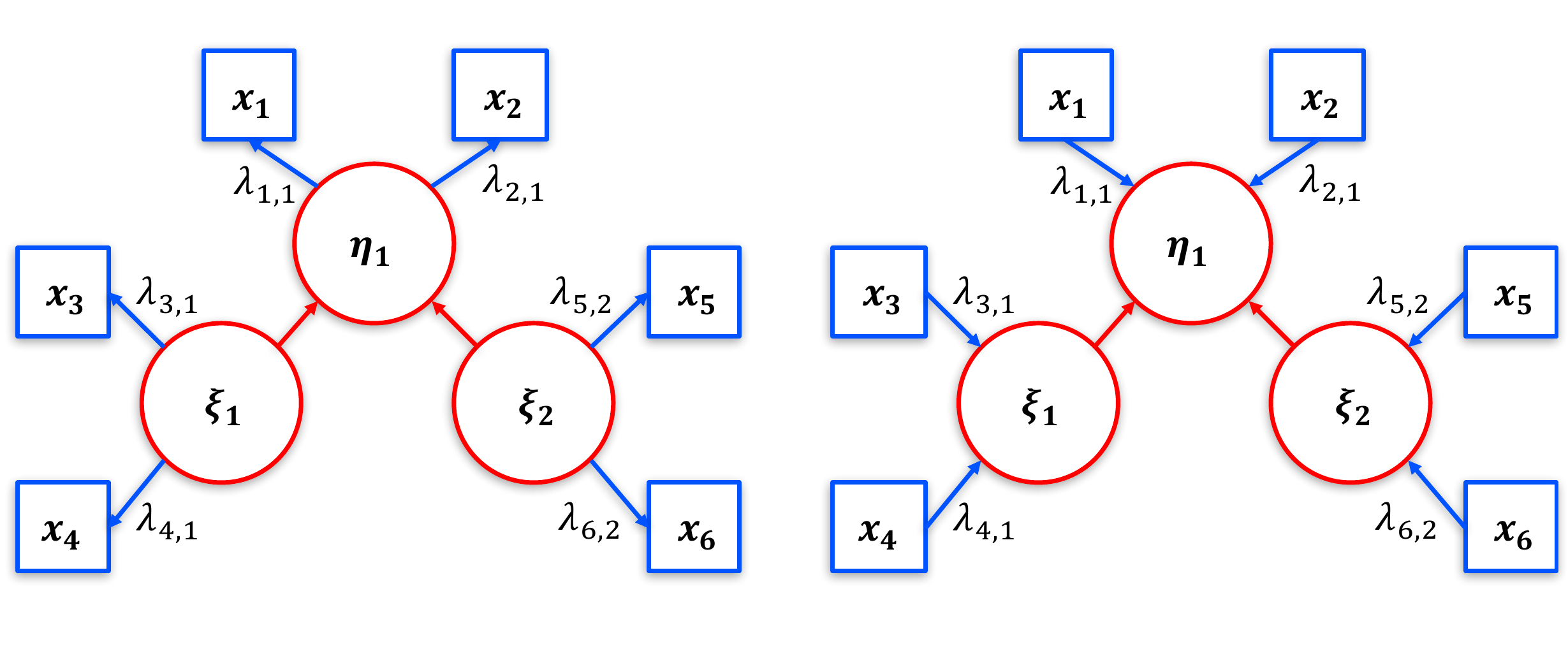}
\caption{Two examples of PLS path model with three LVs and six MVs: reflective measurement models (left) and formative measurement models (right)}\label{fig:3}
\end{figure}

Conversely, in the formative way each MV is supposed "\textit{forming}" its LV, i.e., the observed variables are considered as the cause of the latent construct. Formally, in the case of exogenous latent construct the model can be written as
\begin{equation}\label{eq:4}
    \mathbf{\Xi}=\mathbf{X\Lambda}_H+\mathbf{T},
\end{equation}
whereas, in the case of endogenous latent construct the model can be written as
\begin{equation}\label{eq:5}
    \mathbf{H}=\mathbf{X\Lambda}_L+\mathbf{\Delta},
\end{equation}
where $\mathbf{T}$ and $\mathbf{\Delta}$ are, respectively, the $n\times H$ and $n\times L$ errors matrices with element $\tau_{i,h}$ and $\delta_{i,l}$, under hypothesis of zero mean and is uncorrelated with $x_{i,j}$. Then, the formative way implies that each MV is related to its LV by a multiple regression model with coefficients $\lambda$'s.\\
\underline{\textbf{EXAMPLE 2}}. In Figure \ref{fig:3} are shown two examples of PLS-SEM with three latent constructs ($\eta_1$, $\xi_1$, and $\xi_2$) and six observed variables ($x_1$, $x_2$, $x_3$, $x_4$, $x_5$, and $x_6$). In particular, there are two exogenous LVs ($\xi_1$ and $\xi_2$) and one endogenous LV ($\eta_1$). The MVs are related to their LVs in reflective way (left plot) and formative way (right plot). 
\\Formally, the reflective measurement models represented in the left plot of Figure \ref{fig:3} can be written as
\begin{equation}\label{eq:6}
        \begin{bmatrix} x_1\\ x_2\\ x_3\\ x_4\\ x_5\\ x_6 \end{bmatrix}^T=\begin{bmatrix} \xi_1 & \xi_2 \end{bmatrix}\begin{bmatrix} 0 & 0\\ 0 & 0\\ \lambda_{3,1,} & 0\\ \lambda_{4,1} & 0\\ 0 & \lambda_{5,2}\\ 0 & \lambda_{6,2} \end{bmatrix}_H^T+\begin{bmatrix} \eta_1 \end{bmatrix} \begin{bmatrix} \lambda_{1,1}\\ \lambda_{2,1}\\ 0\\ 0\\ 0\\ 0 \end{bmatrix}_L^T+ \begin{bmatrix} \epsilon_1\\ \epsilon_2\\ \epsilon_3\\ \epsilon_4\\ \epsilon_5\\ \epsilon_6 \end{bmatrix}^T,
\end{equation}
whereas, for the formative measurement models, we can use Equation (\ref{eq:7}) in the case of exogenous LVs, and Equation (\ref{eq:8}) in the case of endogenous LVs.
\begin{equation}\label{eq:7}
        \begin{bmatrix} \xi_1\\ \xi_2 \end{bmatrix}^T=\begin{bmatrix} x_1 & x_2 & x_3 & x_4 & x_5 & x_6 \end{bmatrix}\begin{bmatrix} 0 & 0\\ 0 & 0\\ \lambda_{3,1,} & 0\\ \lambda_{4,1} & 0\\ 0 & \lambda_{5,2}\\ 0 & \lambda_{6,2} \end{bmatrix}_H^T+\begin{bmatrix} \tau_1\\ \tau_2 \end{bmatrix},
\end{equation}
\begin{equation}\label{eq:8}
        \begin{bmatrix} \eta_1 \end{bmatrix}=\begin{bmatrix} x_1 & x_2 & x_3 & x_4 & x_5 & x_6 \end{bmatrix}\begin{bmatrix} \lambda_{1,1}\\ \lambda_{2,1}\\ 0\\ 0\\ 0\\ 0 \end{bmatrix}_L^T+\begin{bmatrix} \delta_1 \end{bmatrix}.
\end{equation}
\section{Partial Least Squares $K$-Means}\label{sec:03}
\subsection{Model and algorithm}
Given the $n\times J$ data matrix $\mathbf{X}$, the $n\times K$ membership matrix $\mathbf{U}$, the $K\times J$ centroids matrix $\mathbf{C}$, the $J\times P$ loadings matrix $\mathbf{\Lambda}$=$[\mathbf{\Lambda}_H, \mathbf{\Lambda}_L]$, and the errors matrices $\mathbf{Z}$ ($n\times L$) and $\mathbf{E}$ ($n\times J$), the partial least squares $K$-means (PLS-SEM-KM) model can be written as follows:
\begin{equation}\label{eq:9}
\begin{split}
        \mathbf{H}&=\mathbf{HB}^T+\mathbf{\Xi\Gamma}^T+\mathbf{Z}\\
        \mathbf{X}&=\mathbf{Y\Lambda}^T+\mathbf{E}=\mathbf{\Xi\Lambda}_H^T+\mathbf{H\Lambda}_L^T+\mathbf{E}\\
        \mathbf{X}&=\mathbf{UC\Lambda\Lambda}^T=\mathbf{UC\Lambda}_H\mathbf{\Lambda}_H^T+\mathbf{UC\Lambda}_L\mathbf{\Lambda}_L^T+\mathbf{E},
\end{split}
\end{equation}
subject to  constraints: $(i)$ $\mathbf{\Lambda}^T\mathbf{\Lambda}=\mathbf{I}$; and $(ii)$ $\mathbf{U}\in \{0,1\}$, $\mathbf{U1}_K=\mathbf{1}_n$. Thus, the PLS-SEM-KM model includes the PLS and the clustering equations (i.e., $\mathbf{X}=\mathbf{UC}$ and then, $\mathbf{Y}=\mathbf{X\Lambda}$ becomes $\mathbf{Y}=\mathbf{UC\Lambda}$). In fact, the third set of equations is the reduced $K$-means model (De Soete and Carroll, 1994). The simultaneous estimation of the three sets of equations will produce the estimation of the presupposed SEM describing relations among variables and the corresponding best partitioning of units.

When applying PLS-SEM-KM, the number of groups $K$ is unknown and the identification of an appropriate number of clusters is not a straightforward task. Several statistical criteria have been proposed. In this paper we use the gap method discussed in Tibshirani et al. (2001) for estimating the number of clusters, i.e., a $pseudo$-F designed to be applicable to virtually any clustering method. In particular, the preliminary step of the PLS-SEM-KM algorithm consists in the selection of the optimal number of classes, according to the maximum level of the $pseudo$-F function computed on the entire data set. The next steps are shown follows.

Given: the $n\times J$ standardized data matrix $\mathbf{X}$; the $J\times P$ design matrix of the measurement model $\mathbf{D}_\Lambda$, with binary elements equal to 1 if a MV is associated to a LV and 0 otherwise; the $P\times P$ path design matrix of the structural model $\mathbf{D}_B$, with binary elements equal to 1 if a latent exogenous or endogenous variable explains a latent endogenous variable and 0 otherwise. Note that the matrix $\mathbf{D}_B$ is symmetrized.
\begin{algorithm}[H]
  \footnotesize
\begin{algorithmic}[1]
\STATE Fix $K$, initialize $\mathbf{\Lambda}=\mathbf{D}_{\Lambda}$ and random memberships matrix $\mathbf{U}$;\\ $\omega=10^{-12}$, iter=0, maxiter=300;
\STATE Compute centers $\mathbf{C}=(\mathbf{U}^T\mathbf{U})^{-1}\mathbf{U}^T\mathbf{X}$ and latent scores $\mathbf{Y}=\mathbf{UC\Lambda}$;\\
\STATE iter=iter+1;\\ 
\vspace{0.2cm}
\fbox{\textbf{Inner approximation}}
\vspace{0.2cm}
\STATE Estimate covariance matrix $\mathbf{\Sigma}_Y=n^{-1}\mathbf{Y}^T\mathbf{J}\mathbf{Y}$ (with $\mathbf{J}=\mathbf{I}n^{-1}\mathbf{1}\mathbf{1}^T$);
\STATE Compute inner weights $\mathbf{W}=\mathbf{D}_B\otimes \mathbf{\Sigma}_{Y}$  ;
\STATE Estimate new scores $\mathbf{Y}_{W}=\mathbf{Y}\mathbf{W}$;\\ 
\vspace{0.2cm}
\fbox{\textbf{Outer approximation}}
\vspace{0.2cm}
\STATE Update $\mathbf{\Lambda}\rightarrow \mathbf{\Lambda}_n=\mathbf{C}^T\mathbf{U}^T\mathbf{Y}_W(\mathbf{Y}_W^T\mathbf{Y}_W)^{-1}$; \: \: 
\STATE Update $\mathbf{U}\rightarrow \underset{\mathbf{U}}{\operatorname{argmin}} \left \|\mathbf{X}-\mathbf{UC\Lambda}_n\mathbf{\Lambda}_n^T\right\|^2$ subject to $\mathbf{\Lambda}_n^T\mathbf{\Lambda}_n=\mathbf{1}_P$, $\mathbf{U}=\{0,1\}$,\\ $\mathbf{U1}_K=\mathbf{1}_n$;   
\STATE Compute new centers $\mathbf{C}_n=(\mathbf{U}^T\mathbf{U})^{-1}\mathbf{U}^T\mathbf{X}$;\\
\vspace{0.2cm}
\fbox{\textbf{Stopping rule}}
\vspace{0.2cm}
\STATE \textbf{if} $\left \|\mathbf{C\Lambda}-\mathbf{C}_n\mathbf{\Lambda}_n\right\|^2>\omega$ \: \& \:  iter$<$maxiter, $\mathbf{C}=\mathbf{C}_n$, $\mathbf{\Lambda}=\mathbf{\Lambda}_n$, $\mathbf{Y}=\mathbf{U}\mathbf{C}_n\mathbf{\Lambda}_n$;\\
repeat step 3-9;\\
\STATE \textbf{else}\\
exit loop 3-9;
\STATE \textbf{end if}\\
\vspace{0.2cm}
\fbox{\textbf{Path coefficients estimation}}
\vspace{0.2cm}
\FOR{$l=1$ to $L$}
\FOR{$h=1$ to $H$}
\STATE Compute $\mathbf{Y}_h=\mathbf{X\Lambda}_h$  
\STATE Compute $\mathbf{Y}_l=\mathbf{X\Lambda}_l$  
\STATE Compute $\mathbf{\Gamma}=(\mathbf{Y}_{h^*}^T\mathbf{Y}_{h^*})^{-1}\mathbf{Y}_{h^*}^T\mathbf{Y}_{l}$ 
\STATE Compute $\mathbf{B}=(\mathbf{Y}_{l^*}^T\mathbf{Y}_{l^*})^{-1}\mathbf{Y}_{l^*}^T\mathbf{Y}_{l}$  
\ENDFOR 
\ENDFOR 
\end{algorithmic}
\caption{PLS-SEM-KM algorithm}
\end{algorithm}
$\mathbf{Y}_h$ is the $h$-th exogenous latent score and $\mathbf{Y}_l$ is the $l$-th endogenous latent score; The symbol $\otimes$ indicates the element-wise product of two matrices, while $*$ indicates the adjacent latent scores matrix, i.e., the set of latent scores that are related to the $\mathbf{Y}_h$ or $\mathbf{Y}_l$. The PLS-SEM-KM algorithm is a development of the Wold's original algorithm used to the PLS-SEM estimate in Lohmöller (1989) and in Lohmöller (2013).\\
Note that, in the PLS-SEM-KM algorithm centroids matrix $\mathbf{C}$ and the loadings matrix $\mathbf{\Lambda}$ simultaneously converge. It is important to remember that the algorithm, given the constraints on $\mathbf{U}$, can be expected to be rather sensitive to \textit{local optima}. For this reasons, it is recommended the use of some randomly started runs to find the best solution.
\subsection{Local and global fit measures}
In PLS-SEM context, there is not a well-identified global optimization criterion to assess the goodness of the model, since PLS-SEM models are variance-based models strongly oriented to prediction and its validation mainly is focused on the predictive capability.\\
According to the PLS-SEM approach, each part of the model needs to be validated: the measurement model, the structural model and the global model (Vinzi et al., 2010). In particular, PLS-SEM provides different fit indices: the \textit{communality} index for the measurement models, the $R^2$ index for the structural models and the \textit{Goodness of Fit} (GoF) index for the overall model.

Communalities are simply the squared correlations between MVs and the corresponding LV. Then, communalities measure the part of the covariance between a latent variable and its block of observed variables that is common to both. For the $j$-th manifest variable of the $p$-th latent score they are calculated as
\begin{equation}\label{eq:10}
        com(x_{j,p},y_p)=corr^2(x_{j,p},y_p).
\end{equation}
Usually, for each $p$-th block of MVs in the PLS-SEM model, the quality of the entire measurement model is assessed by the mean of the communality indices as in equation (\ref{eq:11}).
\begin{equation}\label{eq:11}
        com_p(x_{j,p},y_p)=J_p^{-1}\sum_{j=1}^{J_p}corr^2(x_{j,p},y_p),
\end{equation}
where $J_p$ is the number of the MVs in the $p$-th block. Note that, for each endogenous LV of the structural model we have an $R^2$ interpreted similarly as in any multiple regression model. Then, $R^2$ indicates the amount of variance in the endogenous latent construct explained by its independent latent variables (Vinzi et al., 2010; Sanchez, 2013).\\
Finally, we have the GoF index that is a $pseudo$ Goodness of Fit measure accounting for the model quality at both the measurement and the structural models. GoF is calculated as the geometric mean of the average communality and the average $R^2$ values. Formally,
\begin{equation}\label{eq:12}
        GoF=(PJ_p)^{-1/2}\left [\sum_{p=1}^P\sum_{j=1}^{J_p}corr^2(x_{j,p},y_p)\right ]^{1/2}L^{-1/2}\left[\sum_{p=1}^P \sum_{l=1}^L R^2(y_p,y_l) \right ]^{1/2},
\end{equation}
where $y_h$ is the $h$-th latent score of the exogenous LVs, $y_l$ is the $l$-th latent score of the endogenous LVs, and $L$ is the number of endogenous LVs. It can be observed that if data are generated according to structural and measurement models (\ref{eq:1}) and (\ref{eq:3}) with zero residuals GoF is equal to 1.\\
Note that GoF can be seen as an index of average prediction for the entire model. For instance, a GoF value of 0.78 could be interpreted as if the prediction power of the model is of 78\%. Moreover, the \textit{penalized} R-squared ($R^{2*}$) index is here proposed for assessing also the classification part in the PLS-SEM-KM model (Equation (\ref{eq:13})): 
\begin{equation}\label{eq:13}
        R^{2*}=\left [\bar{R_L^2}\times \left \|\mathbf{UC\Lambda\Lambda}^T\right\|^2/ \left \|\mathbf{X}\right\|^2 \right ]^{1/2},
\end{equation}
where the first term of the product is the mean of $R^2$ computed for each endogenous latent construct, and the second term is the ratio between the model between groups deviance and the total deviance. In this way, the index is more sensible for the mis-classification events. Also $R^{2*}$ assumes values in $[0, 1]$ interval.
\section{Simulation study}
In this section, we have prepared a simulation study for assessing the performances of the partial least squares $K$-means (PLS-SEM-KM) algorithm through a comparison with finite mixture partial least squares (FIMIX-PLS) proposed by Hahn et al. (2002).
\subsection{Simulation scheme}
\begin{figure}[!h]
\centering
\includegraphics[scale=0.6]{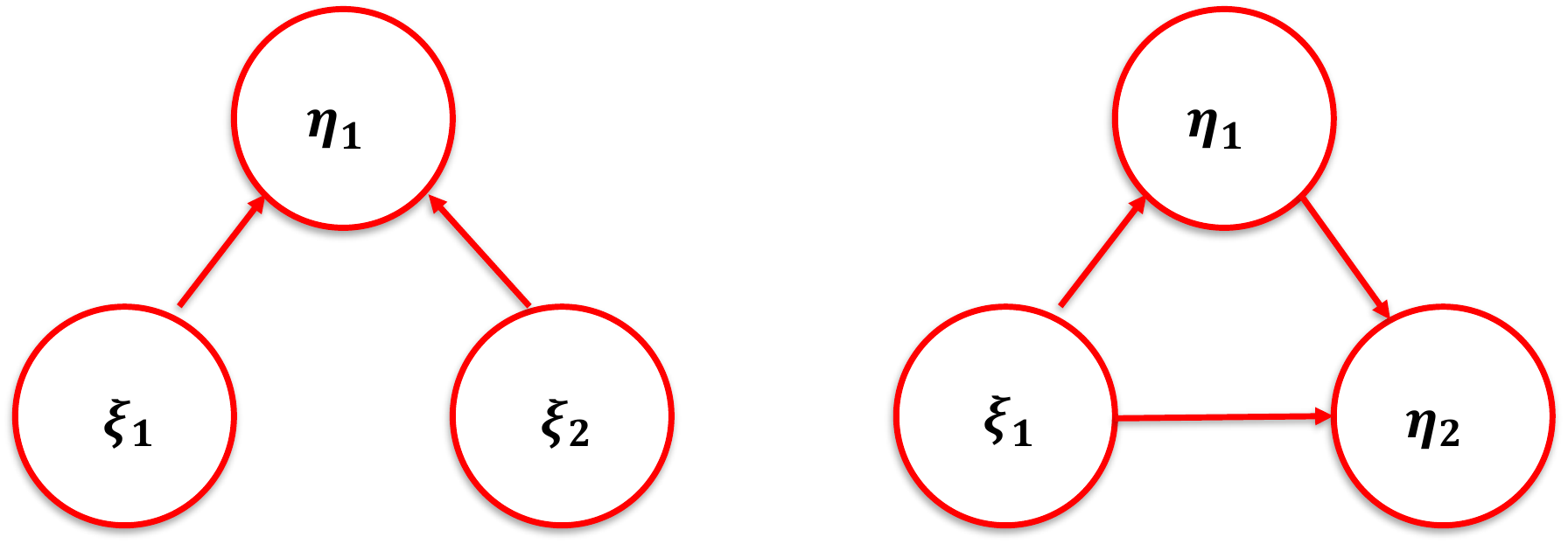}
\caption{Path diagrams of the structural models specified by the simulation scheme}\label{fig:4}
\end{figure}

In order to generate the manifest variables in matrix $\mathbf{X}$, the models shown in equation (\ref{eq:1}) and equation (\ref{eq:3}) have been used. In fact, if we rewrite equation (\ref{eq:1}) as
\begin{equation}\label{eq:14}
\begin{split}
        &\mathbf{H}(\mathbf{I}-\mathbf{B}^T)=\mathbf{\Xi\Gamma}^T+\mathbf{Z}\\
        &\mathbf{H}=\mathbf{\Xi\Gamma}^T(\mathbf{I}-\mathbf{B}^T)^{-1}+\mathbf{Z}(\mathbf{I}-\mathbf{B}^T)^{-1}\\ 
        &\mathbf{H}=\mathbf{\Xi\Gamma}^T(\mathbf{I}-\mathbf{B}^T)^{-1}+error,
\end{split}
\end{equation}
and we include equation (\ref{eq:14}) in equation (\ref{eq:3}), we obtain
\begin{equation}\label{eq:15}
        \mathbf{X}=\mathbf{\Xi\Lambda}_H^T+\mathbf{\Xi\Gamma}^T(\mathbf{I}-\mathbf{B}^T)^{-1}\mathbf{\Lambda}_L^T+\mathbf{E}.
\end{equation}
Then, using the model shown in equation (\ref{eq:15}) we have simulated data matrices formed by different samples of statistical units and 9 MVs ($n\times 9$ data matrices). The 9 generated variables are split in three blocks related to 3 LVs according the path diagram shown in Figure \ref{fig:4}. 
\\The exogenous latent scores matrix $\mathbf{\Xi}$ has been generated by a mixture of $K$ different Normal distributions (2 uncorrelated dimensions for the first path model and 1 dimension for the second path model) to obtain a structure of $K$ groups of units.
The errors matrix $\mathbf{E}$ has been generated by a multivariate Normal distribution (9 uncorrelated dimensions) with means equal to zero (i.e., noise) and standard deviation fixed as: $\sigma=0.30$ (\textit{low error}), $\sigma=0.40$ (\textit{medium error}), $\sigma=0.50$ (\textit{high error}). Finally, all outer and inner loadings (i.e., the matrices $\mathbf{\Lambda}$ and $\mathbf{B}$) have been fixed at 0.85. 
\begin{figure}[!h]
\begin{subfigure}{.33\textwidth}
  \centering
  \includegraphics[width=1\linewidth]{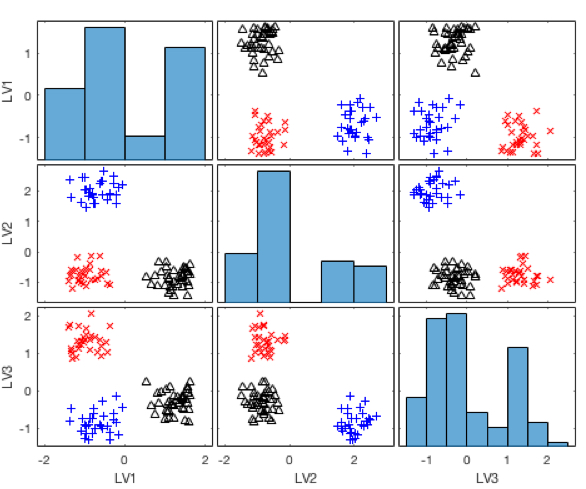}
  \caption{}
  \label{fig:sfig5}
\end{subfigure}%
\begin{subfigure}{.33\textwidth}
  \centering
  \includegraphics[width=1\linewidth]{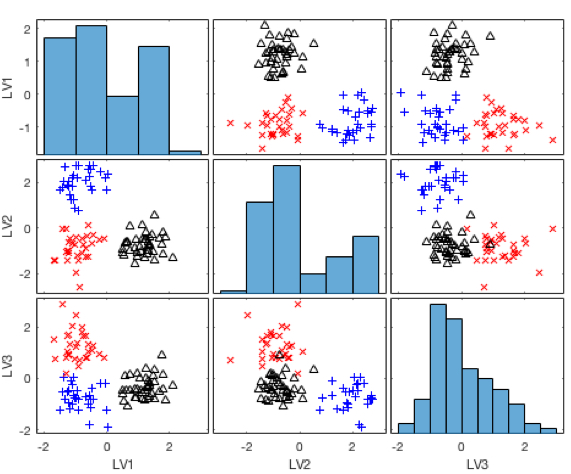}
  \caption{}
  \label{fig:sfig6}
\end{subfigure}
\begin{subfigure}{.33\textwidth}
  \centering
  \includegraphics[width=1\linewidth]{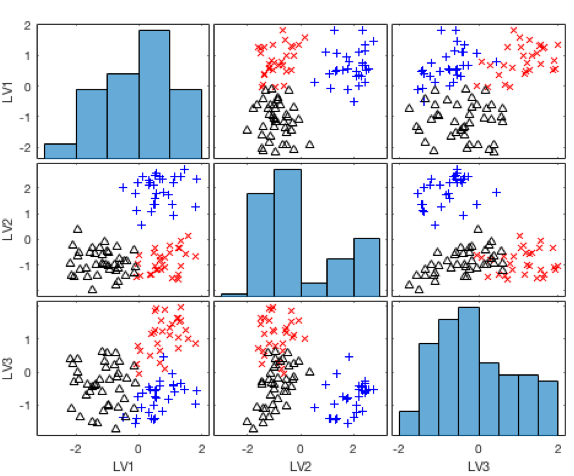}
  \caption{}
  \label{fig:sfig7}
\end{subfigure}
\caption{Scatterplot-matrix of generated data with low, medium and high error.}
\end{figure}

In Figure \ref{fig:sfig5} the scatterplot-matrix of a random generation of the latent scores with low error is shown. Note that LVI is $\xi_1$, LV2 is $\xi_2$ and LV3 is $\eta_1$. The three generated groups with three different colors (30 points blue, 30 points red and 40 points black) and three different symbols ($+$, $\times$, and $\triangle$) are very well-separated and homogenous.\\
Different results are shown with medium error (Figure \ref{fig:sfig6}), where the three groups not well-separated, mostly between the latent scores $\xi_2$ and $\eta_1$.\\
Finally, Figure \ref{fig:sfig7} shows the results obtained with high error that, obviously, correspond to the most confused situation: the groups are not separated and homogeneous, and there is an overlap in all the three couple of latent dimensions.

Moreover, for better investigating the performance of both PLS-SEM-KM and FIMIX-PLS we have realized a simulation study that represents different data constellations that could occur in empirical applications. According to recent simulation studies on PLS segmentation (Becker et al., 2013; Ringle et al., 2014; Schlittgen, 2016), we have selected the following experimental factors:
\begin{itemize}
\item \textit{Number of observations}: \textbf{small sample size} ($n=150$); \textbf{large sample size} ($n=300$).
\item \textit{Number of segments (clusters)}: $K=3$; $K=4$.
\item \textit{Segments size}: \textbf{balanced} (mixture proportion when $K=3$: $p_1=0.33$, $p_2=0.33$, $p_3=0.34$; mixture proportion when $K=4$: $p_1=p_2=p_3=p_4=0.25$); \textbf{unbalanced 1} (mixture proportion when $K=3$: $p_1=0.66$, $p_2=0.17$, $p_3=0.17$; mixture proportion when $K=4$: $p_1=0.40$, $p_2=0.20$, $p_3=0.20$, $p_4=0.20$); \textbf{unbalanced 2} (mixture proportion when $K=3$: $p_1=0.15$, $p_2=0.42$, $p_3=0.43$; mixture proportion when $K=4$: $p_1=0.10$, $p_2=0.30$, $p_3=0.30$, $p_4=0.30$).
\item \textit{Standard deviation of data generation error}: \textbf{low error} ($\sigma=0.30$); \textbf{medium error} ($\sigma=0.40$); \textbf{high error} ($\sigma=0.50$). 
\item \textit{PLS structural model complexity}: \textbf{Model 1} (simple path model shown in left plot of Figure \ref{fig:4}; \textbf{Model 2} (complex path model shown in right plot of Figure \ref{fig:4}.
\end{itemize}
In order to have more stable results, we have randomly generated 100 datasets for each factor level combination. Then, in particular we have $2\times2\times3\times3\times2\times100=7200$ generated datasets.
\subsection{Simulation study results}
We have separated the simulation results in 4 different contexts, each of them with 18 different experimental cases random repeated 100 times. 

In particular, we have context 1: \textit{path model 1 and $K=3$}, context 2: \textit{path model 2 and $K=3$}, context 3: \textit{path model 1 and $K=4$}, and context 4: \textit{path model 2 and $K=4$}. Table \ref{tab:0111} shown the 18 different experimental cases. 
\begin{table*}[h!]
\centering
\caption{Experimental cases list of the simulation study}
\label{tab:0111}
\begin{tabular}{c c c c}
\hline
Exp. Case & Sample Size & Segments size & Error level\\
\hline
1 & small & balanced & low\\
2 & small & balanced & medium\\
3 & small & balanced & high\\
4 & small & unbalanced 1 & low\\
5 & small & unbalanced 1 & medium\\
6 & small & unbalanced 1 & high\\
7 & small & unbalanced 2 & low\\
8 & small & unbalanced 2 & meidum\\
9 & small & unbalanced 2 & high\\
10 & large & balanced & low\\
11 & large & balanced & medium\\
12 & large & balanced & high\\
13 & large & unbalanced 1 & low\\
14 & large & unbalanced 1 & medium\\
15 & large & unbalanced 1 & high\\
16 & large & unbalanced 2 & low\\
17 & large & unbalanced 2 & medium\\
18 & large & unbalanced 2 & high\\
\hline
\end{tabular}
\end{table*}

For evaluating the performance of the models we have used the \textit{penalized} R-squared ($R^{2*}$) index presented in Section 3. In Table \ref{tab:0001} we can see the arithmetic mean of the $R^{2*}$ values obtained for each experimental case of the first and second simulated context by PLS-SEM-KM and FIMIX-PLS, respectively. Similarly, in \ref{tab:0002} we can see the arithmetic mean of each $R^{2*}$ distribution obtained for each experimental case of the third and fourth simulated context by PLS-SEM-KM and FIMIX-PLS, respectively.
\begin{table*}[!h]
\centering
\caption{Mean of the values of $R^{2*}$ obtained by of PLS-SEM-KM and FIMIX-PLS for all experimental cases of the first and second simulated context}
\label{tab:0001}
\begin{tabular}{c | c c | c c c c c c}
& \multicolumn{2}{ c |}{Context 1} & \multicolumn{2}{c}{Context 2}\\
\hline
Exp. Case	 &	PLS-SEM-KM	&	FIMIX-PLS	&	PLS-SEM-KM	&	FIMIX-PLS\\
\hline
1	&	0.981	&	0.954	&	0.718	&	0.577\\
2	&	0.968	&	0.952	&	0.705	&	0.573\\
3	&	0.952	&	0.947	&	0.682	&	0.549\\
4	&	0.980	&	0.751	&	0.513	&	0.378\\
5	&	0.980	&	0.751	&	0.513	&	0.378\\
6	&	0.951	&	0.695	&	0.441	&	0.437\\
7	&	0.978	&	0.714	&	0.437	&	0.375\\
8	&	0.964	&	0.671	&	0.430	&	0.397\\
9	&	0.948	&	0.652	&	0.384	&	0.396\\
10	&	0.981	&	0.976	&	0.722	&	0.578\\
11	&	0.968	&	0.952	&	0.699	&	0.564\\
12	&	0.951	&	0.941	&	0.678	&	0.539\\
13	&	0.980	&	0.743	&	0.488	&	0.397\\
14	&	0.967	&	0.727	&	0.495	&	0.415\\
15	&	0.949	&	0.695	&	0.446	&	0.446\\
16	&	0.979	&	0.708	&	0.437	&	0.364\\
17	&	0.964	&	0.692	&	0.398	&	0.374\\
18	&	0.946	&	0.631	&	0.392	&	0.452\\
\hline
\end{tabular}
\end{table*}
\begin{table*}[!h]
\centering
\caption{Mean of the values of $R^{2*}$ obtained by of PLS-SEM-KM and FIMIX-PLS for all experimental cases of the third and fourth simulated context}
\label{tab:0002}
\begin{tabular}{c | c c | c c c c c c}
& \multicolumn{2}{ c |}{Context 3} & \multicolumn{2}{c}{Context 4}\\
\hline
Exp. Case	 &	PLS-SEM-KM	&	FIMIX-PLS	&	PLS-SEM-KM	&	FIMIX-PLS\\
\hline
1	&	0.975	&	0.926	&	0.703	&	0.623\\
2	&	0.954	&	0.887	&	0.659	&	0.561\\
3	&	0.925	&	0.889	&	0.603	&	0.489\\
4	&	0.963	&	0.828	&	0.694	&	0.564\\
5	&	0.963	&	0.828	&	0.694	&	0.564\\
6	&	0.829	&	0.795	&	0.601	&	0.455\\
7	&	0.808	&	0.648	&	0.643	&	0.465\\
8	&	0.705	&	0.653	&	0.648	&	0.419\\
9	&	0.630	&	0.627	&	0.597	&	0.368\\
10	&	0.975	&	0.938	&	0.684	&	0.645\\
11	&	0.955	&	0.905	&	0.659	&	0.572\\
12	&	0.928	&	0.864	&	0.600	&	0.485\\
13	&	0.977	&	0.862	&	0.694	&	0.576\\
14	&	0.932	&	0.805	&	0.661	&	0.523\\
15	&	0.867	&	0.808	&	0.607	&	0.451\\
16	&	0.843	&	0.650	&	0.668	&	0.449\\
17	&	0.691	&	0.635	&	0.648	&	0.406\\
18	&	0.617	&	0.631	&	0.600	&	0.363\\
\hline
\end{tabular}
\end{table*}
\\

Tables \ref{tab:0001} and \ref{tab:0002} show that the results obtained by PLS-SEM-KM are in almost all cases better than model FIMIX-PLS. In Context 1, where the path model 1 has been considered, the difference between PLS-SEM-KM and FIMIX-PLS are more relevant for cases from 4 to 9 (from 22\% to 29\% better)  and for cases from 13 to 18 (from 23\% to 31\% better), corresponding to the unbalanced cases. In the Context 2 differences are still in favor of the PLS-SEM-KM, but with a less relevant magnitude (no more than 14\%), this time in the balanced cases.

Moreover, the results show also that $R^{2*}$ index for both PLS-SEM-KM and FIMIX-PLS reduces with the increase of the number of segments. This is expected because the probability of misclassification increases. Note that in FIMIX-PLS, as such as in other segmentation models, the correct identification of the number of clusters (segments) is not easy when the number of segments increases. This because FIMIX-PLS follows a mixture regression concept that allows the estimation of separate linear regression functions, and in this way the number of parameters exponentially increases when the number of segments increase, and the usual criteria based on likelihood function, such as AIC and BIC become not very reliable (Bulteel et al., 2013).

Furthermore, it is useful recall that we have generated data from normal mixture model; thus FIMIX-PLS is advantaged since the data for the simulation study are generated according to the FIMIX-PLS hypotheses, by assuming that each endogenous latent variable $\eta_l$ is distributed as a finite mixture of conditional multivariate normal densities (Ringle et al., 2010). Conversely, in PLS-SEM-KM there are not particular assumption on the distribution of data. 

In order to understand the performance of PLS-SEM-KM algorithm we have also studied the presence of \textit{local minima} and situations of \textit{overfitting}. Then, once established that there are cases where the \textit{Adjusted Rand Index} (ARI) is lower than 1 (i.e., the real partition is not identified), it is useful to analyze the single case where the real partition has not been found by PLS-SEM-KM algorithm. 

Table \ref{tab:02} shows the performance of the model in terms of clustering capability for 100 randomly chosen experimental conditions. The second column of the table shows the percentage of times the \textit{gap method}, discussed in section \ref{sec:03}, identifies the real number of clusters ($K=3$ or $K=4$). The third column shows the percentage of times the algorithm finds the true partition; the fourth column shows the percentage of not optimal \textit{local minima}; and finally, the fifth column displays the percentage of \textit{overfitting} in the estimation of the model. 
\begin{table*}[!h]
\centering
\caption{Performance of the PLS-SEM-KM algorithm using a single random start in the three different error levels for 100 randomly chosen experimental conditions (percentage values)}
\label{tab:02}
\begin{tabular}{l c c c c}
\hline
Sd. Error & optimal $K$ & model is $true$ & Local minima & Overfitting\\
\hline
$\sigma=0.30$ & 100.00 & 99.90 & 0.00 &	0.10\\
$\sigma=0.40$ & 100.00 & 88.00 & 7.40 &	4.60\\
$\sigma=0.50$ & 100.00 & 72.60 & 17.20 & 10.20\\ 
\hline
\end{tabular}
\end{table*}

Observing these results, we can try to reduce the number of local minima by increasing the number of initial random starts. In these cases, the use of 15 random starts usually suffices, when the error is not very high ($\sigma=0.30$), and the groups structure is not masked. Indeed, the algorithm finds the optimal solution in 99.90\%; while there are 0.10\% of \textit{overfitting} cases, only, and any \textit{local minima}. However, in the cases where the groups structure is masked as in the case of medium and high level of error, the algorithm cannot completely eliminates the number of \textit{local minima}. In these two cases the algorithm finds the optimal solution in 88\% and 72.60\% of cases, respectively. Thus, it is advisable to increase initial random starts when the clustering of the data is not clear.
\section{Application on real data}
In this section an application on real data of the partial least squares $K$-Means model is presented. For this application the European Consumer Satisfaction Index (ECSI) has been considered analyizing the ECSI approach in mobile phone industry (Bayol et al., 2000; Tenenhaus et al., 2005).  
\subsection{European Consumer Satisfaction Index}
ECSI is an economic indicator that measures the customer satisfaction. It is an adaptation of the Swedish Customer Satisfaction Barometer (Fornell, 1992; Fornell et al., 1996; Hackl and Westlund, 2000) and is compatible with the American Customer Satisfaction Index (ACSI). A model has been derived specifically for the ECSI. In the complete model, there are seven interrelated latent variables as described below. 
\begin{itemize}
    \item \textbf{Image}: refers to the brand name and the kind of associations customers get from the product/brand/company. It is expected that image will have a positive effect on customer satisfaction and loyalty. In addition, image is also expected to have a direct effect on expectations.
    \item \textbf{Expectations}: is the information based on, not actual consumption experienced but, accumulated information about quality from outside sources, such as advertising, word of mouth, and general media.
    \item \textbf{Perceived Quality}: comprises product quality (hardware) and service quality (soft-ware/human-ware). Perceived product quality is the evaluation of recent consumption experience of products. Perceived service quality is the evaluation of recent consumption experience of associated services like customer service, conditions of product display, range of services and products, etc. Perceived quality is expected to affect satisfaction.
    \item \textbf{Perceived Value}: is the perceived level of product quality relative to the price paid of the "value for the money" aspect of the customer experience.
    \item \textbf{Satisfaction}: is defined as an overall evaluation of a firm’s post-purchase performance or utilization of a service. 
    \item \textbf{Complaints}: implies the complaints of customers.
    \item \textbf{Loyalty}: refers to the intention repurchase and price tolerance of customers. It is the ultimate dependent variable in the model and it is expected that the better image and higher customer satisfaction should increase customer loyalty.
\end{itemize}
Moreover, a set of observed variables is associated with each latent variables. The entire model is important for determining the main goal variable, being Customer Satisfaction Index (CSI).
\subsection{ECSI model for mobile phone example}
\begin{figure}[!h]
\centering
\includegraphics[scale=0.37]{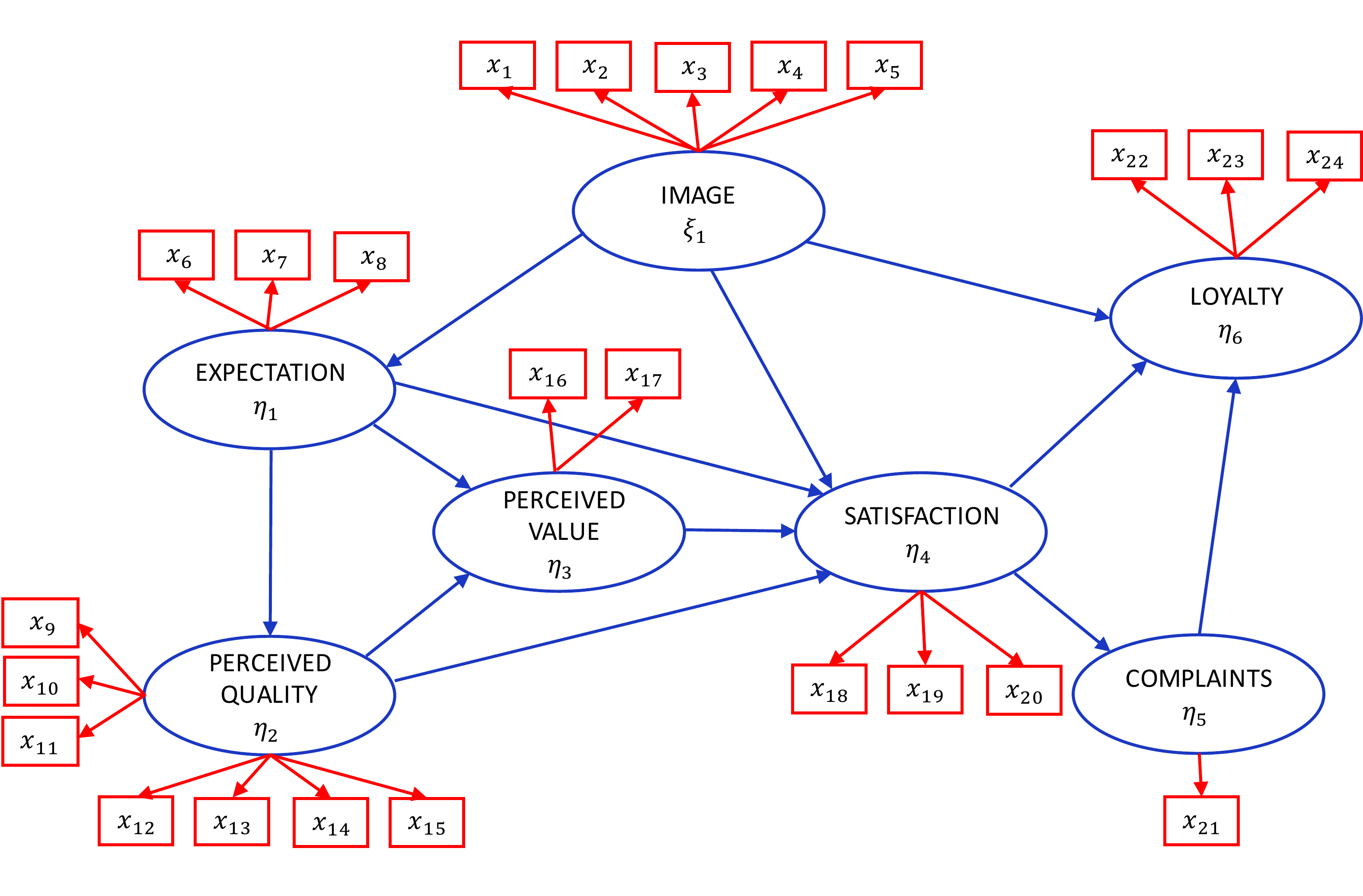}
\caption{ECSI model for the mobile phone industry}\label{fig:11}
\end{figure}

The dataset consists in 24 observed variables that represent the answers of 250 consumers of a mobile phone provider. The original items $v_{j,p}$, scaled from 1 to 10, are transformed into new normalized variables (scaled from 0 to 100) as following:
\begin{equation}\label{eq:17}
        x_{j,p}=\frac{100}{9}(v_{j,p}-1) \in [0, 100]
\end{equation}
The missing data for a variable have been replaced by the mean of the variable. In Figure \ref{fig:11} is represented the complete ECSI model for the mobile phone industry.

The MVs included in dataset are the following:\\
$\mathbf{x_{1}}$: It can be trusted in what it says and does\\
$\mathbf{x_{2}}$: It is stable and firmly established\\
$\mathbf{x_{3}}$: It has a social contribution for the society\\
$\mathbf{x_{4}}$: It is concerned with customers\\
$\mathbf{x_{5}}$: It is innovative and forward looking\\
$\mathbf{x_{6}}$: Expectations for the overall quality of "your mobile phone provider" at the moment you became customer of this provider\\
$\mathbf{x_{7}}$: Expectations for "your mobile phone provider" to provide products and services to meet your personal need\\
$\mathbf{x_{8}}$: How often did you expect that things could go wrong at "your mobile phone provider"\\
$\mathbf{x_{9}}$: Overall perceived quality\\
$\mathbf{x_{10}}$: Technical quality of the network\\
$\mathbf{x_{11}}$: Customer service and personal advice offered\\
$\mathbf{x_{12}}$: Quality of the services you use\\
$\mathbf{x_{13}}$: Range of services and products offered\\
$\mathbf{x_{14}}$: Reliability and accuracy of the products and services provided\\
$\mathbf{x_{15}}$: Clarity and transparency of information provided\\
$\mathbf{x_{16}}$: Given the quality of the products and services offered by "your mobile phone provider" how would you rate the fees and prices that you pay for them?\\
$\mathbf{x_{17}}$: Given the fees and prices that you pay for "your mobile phone provider" how would you rate the quality of the products and services offered by "your mobile phone provider"?\\
$\mathbf{x_{18}}$: Overall satisfaction\\
$\mathbf{x_{19}}$: Fulfillment of expectations\\
$\mathbf{x_{20}}$: How well do you think "your mobile phone provider" compares with your ideal mobile phone provider?\\
$\mathbf{x_{21}}$: \underline{You complained about "your mobile phone provider" last year}.\\
How well, or poorly, was your most recent complaint handled; or\\ 
\underline{You did not complain about "your mobile phone provider" last year}.\\
Imagine you have to complain to "your mobile phone provider" because of a bad quality of service or product. To what extent do you think that "your mobile phone provider" will care about your complaint?\\
$\mathbf{x_{22}}$: If you would need to choose a new mobile phone provider how likely is it that you would choose "your provider" again?\\
$\mathbf{x_{23}}$: Let us now suppose that other mobile phone providers decide to lower their fees and prices, but "your mobile phone provider" stays at the same level as today. At which level of difference (in \%) would you choose another mobile phone provider?\\
$\mathbf{x_{24}}$: If a friend or colleague asks you for advice, how likely is it that you would recommend "your mobile phone provider"?
\subsection{Results}
\begin{figure}[!h]
\centering
\includegraphics[scale=0.05]{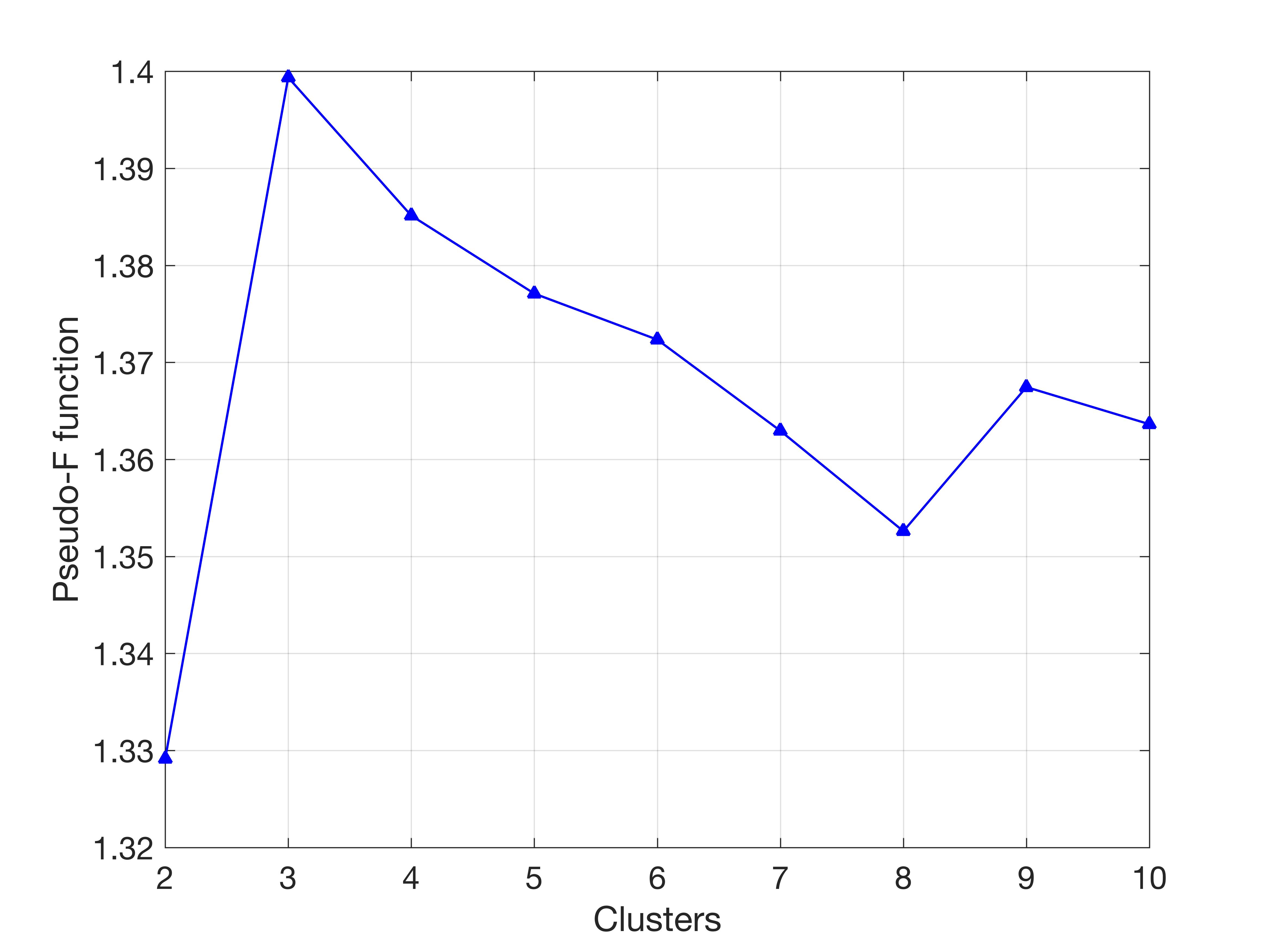}
\caption{$Pseudo$-F function obtained via gap method on the PLS scores from 2 to 10 clusters}\label{fig:12}
\end{figure}

The first step of the analysis consists in the choice of the number of clusters $K$. The \textit{gap method} proposed by Tibshirani at al. (2001) suggests $K=3$, with the corresponding value of $pseudo$-F equal to 1.3994 as shown in Figure \ref{fig:12}.

After the number of clusters has been chosen, the PLS-SEM-KM algorithm can be applied. Table \ref{tab:03} includes the loadings obtained by PLS-SEM-KM (note that variables are standardized).
\begin{table*}[h!]
\centering
\tiny
\caption{Loadings matrix $\mathbf{\Lambda}$ obtained by the PLS-SEM-KM model}
\label{tab:03}
\begin{tabular}{l c c c c c c c}
\hline
MVs & $\xi_1$ &	$\eta_1$ & $\eta_2$ & $\eta_3$ & $\eta_4$ &	$\eta_5$ & $\eta_6$\\
\hline
$x_1$ & 0.449 & 0 & 0 & 0 & 0 & 0 & 0\\
$x_2$ &	0.398 & 0 &	0 &	0 &	0 &	0 &	0\\
$x_3$ &	0.355 &	0 &	0 &	0 &	0 &	0 &	0\\
$x_4$ &	0.528 &	0 &	0 &	0 &	0 &	0 &	0\\
$x_5$ &	0.486 &	0 &	0 &	0 &	0 &	0 &	0\\
$x_6$ &	0 &	0.615 &	0 &	0 &	0 &	0 &	0\\
$x_7$ &	0 &	0.607 &	0 &	0 &	0 &	0 &	0\\
$x_8$ &	0 &	0.503 &	0 &	0 &	0 &	0 &	0\\
$x_9$ &	0 &	0 &	0.419 &	0 &	0 &	0 &	0\\
$x_{10}$ & 0 & 0 & 0.284 & 0 & 0 & 0 & 0\\
$x_{11}$ & 0 & 0 & 0.399 & 0 & 0 & 0 & 0\\
$x_{12}$ & 0 & 0 & 0.377 & 0 & 0 & 0 & 0\\
$x_{13}$ & 0 & 0 & 0.375 & 0 & 0 & 0 & 0\\
$x_{14}$ &	0 &	0 &	0.381 &	0 &	0 &	0 &	0\\
$x_{15}$ &	0 & 0 &	0.397 &	0 &	0 &	0 &	0\\
$x_{16}$ &	0 & 0 &	0 &	0.624 &	0 &	0 &	0\\
$x_{17}$ &	0 &	0 &	0 &	0.781 &	0 &	0 &	0\\
$x_{18}$ &	0 &	0 &	0 &	0 &	0.558 &	0 &	0\\
$x_{19}$ &	0 &	0 &	0 &	0 &	0.563 &	0 &	0\\
$x_{20}$ &	0 &	0 &	0 &	0 &	0.609 &	0 &	0\\
$x_{21}$ &	0 &	0 &	0 &	0 &	0 &	1.000 &	0\\
$x_{22}$ &	0 &	0 &	0 &	0 &	0 &	0 &	0.585\\
$x_{23}$ &	0 &	0 &	0 &	0 &	0 &	0 &	0.099\\
$x_{24}$ &	0 &	0 &	0 &	0 &	0 &	0 & 0.805\\
\hline
\end{tabular}
\end{table*}

Once analyzed the measurement models (i.e., \textit{outer model}), the structural model (i.e., \textit{inner model}) has been considered. Table \ref{tab:04} shows the $P\times P$ path coefficients matrix (note that this matrix includes endogenous and exogenous path coefficients, i.e., $\mathbf{B}$ and $\mathbf{\Gamma}$) estimated by the model (the latent scores have been standardized).
\begin{table*}[h!]
\centering
\tiny
\caption{Path coefficients matrix obtained by the PLS-SEM-KM model}
\label{tab:04}
\begin{tabular}{l c c c c c c c}
\hline
& $\xi_1$ &	$\eta_1$ & $\eta_2$ & $\eta_3$ & $\eta_4$ &	$\eta_5$ & $\eta_6$\\
\hline
$\xi_1$ & 0 & 0.507 & 0 & 0 & 0.177 & 0 & 0.201\\
$\eta_1$ & 0 &	0 &	0.554 &	0.048 &	0.071 &	0 &	0\\
$\eta_2$ & 0 &	0 &	0 &	0.557 &	0.509 &	0 &	0\\
$\eta_3$ & 0 &	0 &	0 &	0 &	0.191 &	0 &	0\\
$\eta_4$ &	0 &	0 &	0 &	0 &	0 &	0.523 &	0.479\\
$\eta_5$ & 0 &	0 &	0 &	0 &	0 &	0 &	0.067\\
$\eta_6$ & 0 & 0 & 0 & 0 & 0 & 0 & 0\\
\hline
\end{tabular}
\end{table*}

Table \ref{tab:04} shows that the \textit{Image} construct has a positive relationship with the all its endogenous LVs, though it has a stronger effect on the \textit{Expectations} construct (0.51) than \textit{Satisfaction} (0.18) and \textit{Loyalty} (0.20). The \textit{Expectations} construct has a significant effect on the \textit{Perceived Quality} only (0.55), while it has very low effect on the \textit{Perceived Value} (0.05) and \textit{Satisfaction} (0.07). The \textit{Perceived Quality} block has effect on \textit{Perceived Value} and \textit{Satisfaction} (correlation values equal to 0.56 and 0.51, respectively). The \textit{Perceived Value} construct has an effect equal to 0.19 on the \textit{Satisfaction}, which has an effect equal to 0.52 on the \textit{Complaints}. Finally, the \textit{Complaints} construct has effect on the \textit{Loyalty} only, with a correlation level equal to 0.07. 

Now, we show the results obtained on the model assessment. In particular we have obtained the communality average equal to 0.5916, the $R^2$ average equal to 0.3872, and the \textit{Goodness of Fit} equal to 0.478. In Table \ref{tab:05} we can see fit measures computed on each latent construct.
\begin{table*}[h!]
\centering
\tiny
\caption{Fit measures computed on each block of MVs}
\label{tab:05}
\begin{tabular}{l c c c}
\hline
LVs & Communality	&	R-Squared	&	Cronbach'alpha	\\
\hline
Image &	0.200 &		0.000 &		0.997\\	
Expectations &	0.333 &		0.257 &		0.998\\	
PercQuality &	0.143 &		0.307 &		0.999\\	
PercValue &	0.500 &	0.342 &		0.990\\	
Satisfaction &	0.333 &		0.677	 &	1.000\\	
Complaints &	1.000 	&	0.274 &		1.000\\	
Loyalty &	0.333 &		0.454 &		0.919\\	
\hline
\end{tabular}
\end{table*}
\\From Table \ref{tab:05} we can say that the PLS-SEM-KM model shows good performances in terms of unidimensionality of the reflective blocks of the variables, communalities, $R^2$ (on endogenous latent variables only) and GoF.
\\

The last step of the analysis is the description of the groups defined by PLS-SEM-KM model. Table \ref{tab:06} shows the summary statistics of the three found groups computed on the seven normalized latent scores. 
\begin{table*}[h!]
\centering
\tiny
\caption{Summary statistics of the three groups of mobile phone customers}
\label{tab:06}
\begin{tabular}{l c c c c c c c}
\hline
& \multicolumn{7}{c}{Group 1 ($n=92$)}\\
 & $\xi_1$ &	$\eta_1$ & $\eta_2$ & $\eta_3$ & $\eta_4$ &	$\eta_5$ & $\eta_6$\\
\hline
$Min$ & 0.460 & 0.180 & 0.660 & 0.000 & 0.537 & 0.000 & 0.019\\
$Q1$ & 0.722 & 0.652 &	0.775 &	0.688 &	0.710 &	0.778 &	0.824\\
$Median$ &0.802 & 0.773 & 0.837 & 0.778 & 0.787 & 0.889 & 0.898\\
$Mean$ & 0.796 & 0.752 & 0.840 & 0.763 & 0.794 & 0.832 & 0.862\\
$Q3$ &	0.861 &	0.849 &	0.905 &	0.878 &	0.875 &	1.000 &	0.956\\
$Max$ &	1.000 &	1.000 &	1.000 &	1.000 & 1.000 &	1.000 &	1.000\\
\hline
\hline
& \multicolumn{7}{c}{Group 2 ($n=112$)}\\
 & $\xi_1$ &	$\eta_1$ & $\eta_2$ & $\eta_3$ & $\eta_4$ &	$\eta_5$ & $\eta_6$\\
\hline
$Min$ &	0.225 &	0.145 &	0.483 &	0.000 &	0.273 &	0.000 &	0.190\\
$Q1$ &	0.541 &	0.481 &	0.594 &	0.511 &	0.526 &	0.556 &	0.594\\
$Median$ &	0.600 &	0.584 &	0.648 &	0.622 &	0.599 &	0.667 &	0.698\\
$Mean$ &	0.607 &	0.584 &	0.643 &	0.591 &	0.589 &	0.638 &	0.696\\
$Q3$ &	0.681 &	0.664 &	0.687 &	0.667 &	0.647 &	0.778 &	0.804\\
$Max$ &	0.845 &	1.000 &	0.831 &	0.889 &	1.000 &	1.000 &	1.000\\
\hline
\hline
& \multicolumn{7}{c}{Group 3 ($n=46$)}\\
 & $\xi_1$ &	$\eta_1$ & $\eta_2$ & $\eta_3$ & $\eta_4$ &	$\eta_5$ & $\eta_6$\\
\hline
$Min$ &	0.000 &	0.000 &  0.000 & 0.000 & 0.000 & 0.000 & 0.000\\
$Q1$ &	0.306 &	0.359 &	0.284 &	0.333 &	0.272 &	0.333 &	0.263\\
$Median$ &	0.440 &	0.497 &	0.414 &	0.444 &	0.353 &	0.444 &	0.467\\
$Mean$ &	0.392 &	0.471 &	0.398 &	0.423 &	0.345 &	0.447 &	0.460\\
$Q3$ &	0.494 &	0.599 &	0.486 &	0.556 &	0.445 &	0.667 &	0.626\\
$Max$ &	0.676 &	0.820 &	0.704 &	1.000 &	0.691 &	1.000 &	1.000\\
\hline
\end{tabular}
\end{table*}

The first group, formed by 92 observations, indicates a \textit{highly satisfied} profile of customers (central values around the 0.8); the second group, formed by 112 observations, indicates a \textit{medially satisfied} profile of customers (central values around 0.6); the third group, formed by 46 observations, indicates a \textit{lowly satisfied} profile of customers (central values around the 0.4).

\section{Conclusions}
In a wide range of applications, the assumption that data are collected from a single homogeneous population, is often unrealistic, and the identification of different groups (clusters) of observations constitutes a critical issue in many fields. 

This work is focused on the structural equation modeling (SEM) in the PLS-SEM context, (i.e., SEM estimated via partial least squares (PLS) method), when the data are heterogeneous and tend to form clustering structures. We know that the traditional approach to clustering in SEM consists in estimating separate models for each cluster, where the partition is \textit{a priori} specified by the researcher or obtained via clustering methods. Conversely, the partial least squares $K$-means (PLS-SEM-KM) approach, provides a single model that guarantees the best partition of objects represented by the best causal relationship in the reduced latent space. 

The simulation study has highlighted a good reliability of the model, which guarantees good results in different experimental cases, when the data have a clustering structure; conversely, the sequential approach to use PLS- SEM followed by clustering on the latent variables may fail to identify the correct clusters. The simulation study shows that in almost all experimental cases PLS-SEM-KM achieves better than finite mixture partial least squares (FIMIX-PLS) model proposed by Hahn et al. (2002). Moreover, we recall FIMIX-PLS in the simulation study has been advantaged since it is based on the assumption that each endogenous latent construct is distributed as a finite mixture of multivariate normal densities, and we have generated data from mixtures of normal distributions. However, imposition of a distributional assumption on the endogenous latent variables may prove to be problematic. This criticism gains force when one considers that PLS path modeling is generally preferred to covariance structure analysis (CSA) in circumstances where assumptions of multivariate normality cannot be made (Ringle et al., 2010). Conversely, in PLS-SEM-KM there are not distributional assumptions. Another problem that was found for  FIMIX-PLS, as such as for other segmentation models, is the correct identification of the number of clusters (segments) when it increases since the approach follows a mixture of regressions; concept that needs the estimation of separate linear regression functions. In this way the number of parameters exponentially increases at the increasing of the number of segments, and the usual criteria based on likelihood function, as such as AIC and BIC are not very reliable. In the PLS-SEM-KM algorithm the gap-method proposed by Tibshirani et al., 2001 is used, and the simulation study shows that the real number of clusters is identified in 100\% of cases in all the simulated contexts.     

On the other hand, in the application on real data we can say that PLS-SEM-KM, in the optimal case (i.e., when the causal structure of the model well-represents the partition that characterizes the data), does not particularly modify the results on the structural and measurement models obtained by the simple PLS-SEM as shown in literature (Bayol et al., 2000; Tenenhaus et al., 2005).  

However, in future research could be interesting to evaluate the PLS-SEM-KM performance against the more recent approaches, as prediction oriented segmentation in PLS path models (PLS- POS) proposed by Becker et al. (2013), genetic algorithm segmentation in partial least squares path modeling (PLS-GAS) proposed by Ringle et al., (2013), and particularly segmentation of PLS path models through iterative reweighted regressions (PLS-IRRS) proposed by Schlittgen et al. (2016).

\end{document}